\documentclass[letterpaper,aps,prl,twocolumn,showpacs,preprintnumbers,superscriptaddress]{revtex4}

\usepackage{graphicx}
\usepackage{epsf} 
\usepackage{dcolumn}
\usepackage{bm}
\usepackage{amssymb}
\usepackage{amsmath}

\usepackage{times}

\begin{document}

\title{Scaling of Local Slopes, Conservation Laws
and Anomalous Roughening in Surface Growth}

\author{Juan M. L\'opez}\email{lopez@ifca.unican.es}
\affiliation{Instituto de F\'{\i}sica de Cantabria (IFCA),
CSIC--UC, E-39005 Santander, Spain}
\author{Mario Castro}
\affiliation{Grupo Interdisciplinar de Sistemas Complejos (GISC)
and Grupo de Din´amica No Lineal (DNL), Escuela T\'ecnica Superior
de Ingenier{\'\i}a (ICAI), Universidad Pontificia Comillas,
E-28015 Madrid, Spain}
\author{Rafael Gallego}
\affiliation{Departamento de Matem\'aticas, Universidad de Oviedo,
Campus de Viesques,
E-33203 Gij\'on,
Spain}

\date{\today}

\begin{abstract}
We argue that symmetries and conservation laws greatly restrict the form of the terms
entering the long wavelength description of growth models exhibiting anomalous
roughening. This is exploited to show by dynamic renormalization group arguments that
intrinsic anomalous roughening cannot occur in local growth models. However some
conserved dynamics may display super-roughening if a given type of terms are present.

\end{abstract}


\maketitle

Recent theoretical and experimental studies of self-affine kinetic
roughening have uncovered a rich variety of novel features
\cite{krug-rev}. In particular, the existence of {\em anomalous}
roughening has received much attention. Anomalous roughening refers
to the observation that {\em local} and {\em global} surface
fluctuations may have distinctly different scaling exponents. This
leads to the existence of an independent {\em local} roughness
exponent $\alpha_{loc}$ that characterizes the local interface
fluctuations and differs from the {\em global} roughness exponent
$\alpha$. More precisely, {\em global} fluctuations are measured by
the global interface width, which for a system of total lateral size
$L$ scales according to the Family-Vicsek ansatz \cite{fv} as
\begin{equation}
\label{FV-globalwidth} W(L,t) = t^{\beta} f(L/t^{1/z}),
\end{equation}
where the scaling function $f(u)$ behaves as
\begin{equation}
\label{FV-forf}
f(u) \sim \left\{ \begin{array}{lcl}
     u^{\alpha}     & {\rm if} & u \ll 1\\
     {\rm const.} & {\rm if} & u \gg 1
\end{array}
\right..
\end{equation}
The roughness exponent $\alpha$ and the dynamic exponent $z$
characterize the universality class of the model under study. The
ratio $\beta = \alpha/z$ is the time exponent. In contrast, {\em
local} surface fluctuations are given by either the height-height
correlation function, $G(l,t) = \langle \overline{[h({\bf x}+{\bf
l},t) - h({\bf x},t)]^2}\rangle$, where the average is calculated
over all ${\bf x}$ (overline) and noise (brackets), or the local
width, $w(l,t) = \langle \langle [h({\bf x},t) - \langle h
\rangle_l]^2\rangle_l\rangle^{1/2}$, where $\langle \cdots
\rangle_{l}$ denotes an average over ${\bf x}$ in windows of size
$l$. For growth processes in which an anomalous roughening takes
place these functions scale as $w(l,t) \sim \sqrt{G(l,t)} =
t^{\beta} f_A(l/t^{1/z})$, with an anomalous scaling function
\cite{krug,lopez97} given by
\begin{equation}
\label{f_A}
f_A(u) \sim \left\{ \begin{array}{lcl}
     u^{\alpha_{loc}}     & {\rm if} & u \ll 1 \\
     {\rm const.} & {\rm if} & u \gg 1
\end{array}
\right.
\end{equation}
instead of Eq.(\ref{FV-forf}). The standard self-affine
Family-Vicsek scaling \cite{fv} is then recovered when $\alpha =
\alpha_{loc}$.

This singular phenomenon was first noticed in numerical
simulations of both continuous and discrete models of ideal
molecular beam epitaxial growth
\cite{krug,lopez97,groove,schro,das,kotrla,infrared,ryu,dasgupta}.
Anomalous roughening has later on been reported to occur in growth
models in the presence of disorder \cite{lopez96,ala},
electro-chemical deposition models \cite{mario}, chemical-vapor
deposition \cite{cvd}, etc.

Remarkably, anomalous roughening has also been reported in many
experimental studies including molecular-beam epitaxy of Si/Si(111)
\cite{yang}, sputter-deposition growth of Pt on glass \cite{jef}, Cu
electrodeposition \cite{huo}, growth of Fe-Cr super-lattices
\cite{jacobo}, propagating fracture cracks \cite{fracture},
fluid-flow in porous media \cite{soriano}, tumor growth \cite{bru},
etc.

Nowadays it has become clear \cite{krug,lopez99} that anomalous
kinetic roughening is related to a non-trivial dynamics of the
average surface gradient (local slope), so that $\langle
\overline{(\nabla h)^2}\rangle \sim t^{2\kappa}$. Anomalous
scaling occurs whenever $\kappa > 0$, which leads to the existence
of a local roughness scaling with exponent $\alpha_{loc} = \alpha
- z\kappa$ \cite{lopez99}. Also, it has recently been shown that
the existence of power-law scaling of the correlation functions
({\it i.e.} scale invariance) does not determine a unique dynamic
scaling form of the correlation functions \cite{ramasco}. On the
one hand, there are {\em super-rough} processes, $\alpha > 1$, for
which $\alpha_{loc} = 1$ always. On the other hand, there are {\em
intrinsically} anomalous roughened surfaces, for which the local
roughness $\alpha_{loc} < 1$ is actually an independent exponent
and $\alpha$ may take values larger or smaller than one depending
on the universality class (see \cite{lopez97,ramasco} and
references therein). The existence of intrinsically anomalous
roughened surfaces is a most intriguing observation and leads to
the still open question concerning the basic physical features
(symmetries, form of the nonlinearities, conservation laws,
non-locality, etc) required for anomalous roughening to occur in
surface growth.

In this Letter we show that local models of surface growth driven by
random noise cannot exhibit intrinsic anomalous roughening.
Nevertheless, some models with  conserved dynamics may display
super-roughening if the required terms are present (see below). Our
results are based upon a dynamic renormalization group analysis of
stochastic growth models that include only local interaction terms
(derivatives of the height field) and the deposition noise is
Gaussian and uncorrelated in space and time. This framework is very
broad and includes many universality classes of growth, in
particular stochastic equations used to theoretically describe
molecular beam epitaxy as well as other types of thin-film growth.
Our results imply that disorder and/or non-local effects are
responsible for intrinsic anomalous roughening in experiments in
different systems \cite{yang,jef,huo,jacobo,fracture,soriano}. We
also argue that the anomalous scaling exponents found in numerical
simulations of discrete growth models (see for instance
\cite{schro,kotrla}) must be effective and corresponding to a
non-universal transient regime due to the use of too small and too
short time scales compared with the true asymptotic regime.

\paragraph{Scaling of the local slopes in surface growth.--}
We are interested here in surface growth models with local coupling
among degrees of freedom, {\it i.e}, growth equations that include
only derivatives of the height. In the long wavelength limit, we
consider surface growth in $d+1$ dimensions described by the
Langevin equation
\begin{equation}
\label{langevin} \partial_t h = {\cal G}(\nabla h) + \eta({\bf
x},t),
\end{equation}
where $h({\bf x},t)$ is the height of the interface at substrate
position ${\bf x}$ and time $t$. The function ${\cal G}(\nabla h)$
defines a particular model and incorporates the relevant symmetries
and conservation laws. In particular, invariance under translation
along the growth and substrate directions as well as invariance in
the election of the time origin rule out an explicit dependence of
${\cal G}$ on $h$, ${\bf x}$ and $t$. Growth is driven by an
external noise $\eta({\bf x},t)$, which represents the influx of
particles in deposition processes. The noise is usually considered
to be Gaussian, uncorrelated in space and time, and either
non-conserved, $\langle \eta({\bf x},t) \eta({\bf x}',t')\rangle =
2D\: \delta({\bf x}-{\bf x}') \delta(t-t')$, or conserved, $\langle
\eta_c({\bf x},t) \eta_c({\bf x}',t')\rangle = -2D\: \nabla^2
\delta({\bf x}-{\bf x}') \delta(t-t')$, depending on the model under
consideration.

The local slope fluctuations scale as $\langle \overline{(\nabla h)^2}\rangle
\sim t^{2\kappa}$. When $\kappa
> 0$ the local slope fluctuations become a relevant scale (in the
growth direction) and local scaling behavior is expected to be anomalous, Eq.(\ref{f_A}),
instead of standard self-affine Family-Vicsek. One can exploit this simple observation
\cite{lopez99} to obtain the local roughness exponent $\alpha_{loc} = \alpha - z\kappa$
by analyzing the corresponding time evolution equation for the local derivative $\Upsilon
= \nabla h$
\begin{equation}
\label{slopes} \partial_t \Upsilon = \nabla {\cal G}(\Upsilon)+
\eta_c({\bf x},t),
\end{equation}
where the noise $\eta_c = \nabla \eta$ is now conserved. Assuming
periodic boundary conditions in Eq.(\ref{langevin}) implies
$\overline{\nabla h} = 0$ and one then finds that the global width
of the interface $\Upsilon({\bf x},t)$ is
\begin{equation}
\label{W_U} W_\Upsilon(t) = \langle \overline{[\Upsilon({\bf x},t)
- \overline{\Upsilon({\bf x},t)}]^2}\rangle^{1/2} = \langle
\overline{(\nabla h)^2}\rangle^{1/2},
\end{equation}
which immediately leads to $W_\Upsilon(t) \sim t^{\kappa}$
\cite{lopez99}. Therefore, by studying the scaling behavior of
surface $\Upsilon({\bf x},t)$ one can obtain the anomalous time
growth exponent $\kappa$ and therefore the local roughness
exponent of the surface $h({\bf x},t)$ through the scaling
relation $\alpha_{loc} = \alpha - z\kappa$.

In all respects we can regard the slopes $\Upsilon({\bf x},t)$ as a
growing surface on its own and study its roughening properties, if
any. Specifically, time scale invariance of (\ref{W_U}) implies that
the global width is expected to obey Eqs.(\ref{FV-globalwidth}) and
(\ref{FV-forf}), so that $W_\Upsilon(L,t) \sim t^\kappa$ for times
$t \ll L^{\hat{z}}$, and saturates, $W_\Upsilon(L,t) \sim
L^{\hat{\alpha}}$, for long times $t \gg L^{\hat{z}}$, where
$\hat{\alpha}$ and $\hat{z} = \hat{\alpha}/\kappa$ are the roughness
and dynamic exponents of the interface defined by the local slopes,
respectively. The growth equation for $\Upsilon({\bf x},t)$,
Eq.(\ref{slopes}), may yield an interface $\Upsilon({\bf x},t)$ that
is not rough ($\hat{\alpha} < 0$). However, depending on the model
symmetries, {\it i.e.}, the form of the terms entering ${\cal G}$ in
(\ref{slopes}), the corresponding local slope $\Upsilon({\bf x},t)$
may turn out to be rough ($\hat{\alpha} > 0$), which
straightforwardly implies anomalous scaling behavior of the surface
$h({\bf x},t)$.

We now analyze the dynamic scaling behavior of the surface slopes
described by Eq.(\ref{slopes}). Let us consider the expansion of
${\cal G}(\Upsilon)$ in powers of $\Upsilon$, its derivatives, and
combinations thereof ({\it i.e.} the leading-order gradient
expansion) consistent with all the symmetries of the problem under
study \cite{barabasi}. The corresponding expansion may have several
terms. However, the long wavelength limit (after renormalization) is
governed by the most relevant term. We do not need to know
explicitly the form of the most relevant term in the expansion, but
we just assume that it generically behaves as ${\cal G}(\Upsilon)
\to b^{-n}\,\Upsilon^m$, with $n \geq 0$ and $m \geq 1$, when the
scale transformation ${\bf x} \to b\,{\bf x}$ is applied
\cite{barabasi}. Note that this term does not have necessarily to be
present in that precise form in the expansion, but it can be
generated by renormalization of other terms. To illustrate this,
consider for instance the $\nabla(\nabla h)^3$ nonlinearity that is
known to renormalize to the linear $\nabla^2 h$ term
\cite{4th_order}. In other words, we may not know {\em a priori} the
values of $n$ and $m$, but we can certainly assume that such a
scaling must exist.

The dynamics of the surface slopes, Eq.(\ref{slopes}), is always
conserved by construction. This immediately leads to the
hyperscaling relation $\hat{z} = 2\hat{\alpha} + d + 2$ for the
slope critical exponents, which is an exact outcome from the
non-renormalization of the noise intensity $D$ for systems with
conserved dynamics \cite{sgg,barabasi}. 
This hyperscaling relation is exact, independent of the detailed
form of ${\cal G}$, 
and allows us to directly link the anomalous time exponent to the
dynamics of the local slopes by $2\kappa = 1 - (d+2)/\hat{z}$.

Since we assume that the leading term of the expansion
renormalizes as $G(\Upsilon) \sim b^{-n}\,\Upsilon^m$, critical
behavior can be obtained from dimensional analysis
\cite{barabasi,flory,lopez99}. Applying the self-affine scale
transformation ${\bf x} \to b\,{\bf x}$,  $t \to b^z\,t$, and
$\Upsilon \to b^{\hat{\alpha}}\, \Upsilon$ to Eq.(\ref{slopes}),
we obtain that the relevant terms scale as $(\partial
\Upsilon/\partial t) \to b^{\hat{\alpha}-\hat{z}}\,(\partial
\Upsilon/\partial t)$, $\nabla{\cal G}(\Upsilon) \to
b^{\hat{\alpha}\,m - (n+1)}\,\nabla{\cal G}(\Upsilon)$, and the
conserved noise scales simply as $\eta_c \to
b^{-1-(d+\hat{z})/2}\,\eta_c$, since it does not renormalize.
These three terms are equally relevant after rescaling only if the
following scaling relationships are satisfied,
\begin{equation}
\label{scal-rel-slope} \left\{ \begin{array}{l}
     (m-1)\hat{\alpha} + \hat{z} - (n+1) = 0   \\
     2\hat{\alpha} - \hat{z} + d + 2 = 0,
\end{array}
\right.
\end{equation}
where we again obtain the hyperscaling relation due to the conserved
dynamics of the surface slopes, as expected. These two basic scaling
relations must be satisfied by the critical exponents of the surface
slopes for any growth model with the appropriate (renormalized)
values of $n$ and $m$. From (\ref{scal-rel-slope}) we can obtain
formal expressions for the critical exponents of the local slope,
$\hat{\alpha} = (n-d-1)/(m+1)$ and $\kappa = \hat{\alpha}/\hat{z} =
(n-d-1)/ [2n-d+(d+2)m]$. Although the values of $n$ and $m$ can only
be determined by dynamical renormalization group techniques for each
particular model, we do not need in this Letter to know their
explicit values. For the sake of our argument, we only need to
observe that there must exist constraints on the possible values
that $n$ and $m$ can have if the local slopes $\Upsilon({\bf x},t)$
have to be rough. The conditions $\hat{\alpha} > 0$ and $\kappa > 0$
yield
\begin{equation}
\label{AS-condit}
     n > d+1,
\end{equation}
which severely restrict the form in which the most relevant term
of ${\cal G}(\Upsilon)$ can renormalize in surface models
exhibiting anomalous roughening. Consider for instance $1+1$
anomalous surface growth, the condition (\ref{AS-condit}) implies
that the most relevant contribution in the expansion of the
relaxation term must behave under rescaling as ${\cal G}(\nabla h)
\to b^{-3}(\nabla h)^m$, where $m$ can be any $m \geq 1$. The most
relevant term of this type corresponds to $m=1$, {\it e.g.} the
linear term $\nabla^4 h$. In fact, the linear equation $\partial_t
h = - \nabla^4h + \eta$ is known to exhibit (super-rough)
anomalous scaling. However, even if the $\nabla^4 h$ is not
present (or not relevant), a most relevant nonlinear term might
exist and satisfy Eq.(\ref{AS-condit}),
yielding $\hat{\alpha}> 0$ and $\kappa > 0$,
and therefore anomalous scaling.

In the following sections we analyze separately conserved and
non-conserved surface dynamics and show that the requirement of
local dynamics together with invariance of the growth equations
under some symmetry transformations results in the impossibility
of intrinsic anomalous scaling. Certain types of conserved
dynamics, however, may exhibit super-roughening scaling
properties.

\paragraph{Conserved dynamics.--} If the surface height $h({\bf
x},t)$ is locally conserved the growth equation (\ref{langevin})
must obey the continuity equation
\begin{equation}
\label{conserv}
\partial_t h = - \nabla\cdot
{\bf j} + \eta({\bf x},t),
\end{equation}
where ${\cal G}(\nabla h)= -\nabla\cdot {\bf j}$. The term ${\bf j}(\nabla h)= {\bf
j}_{eq} + {\bf j}_{neq}$ includes equilibrium and nonequilibrium contributions to the
current. Invariance under the choice of the height origin implies that the current ${\bf
j}$ cannot depend explicitly on $h$ \cite{barabasi}. This type of equation has attracted
much attention in recent years because of its relevance in the continuum theory of
molecular beam epitaxial growth. We can follow the same procedure as above to evaluate
the scaling behavior of the surface $h({\bf x},t)$ from Eq.(\ref{langevin}) when ${\cal
G}(\nabla h) = -\nabla\cdot {\bf j}$. Conservation implies that the noise scales as $\eta
\to b^{-(d+z)/2}\, \eta$. For the velocity term we have $(\partial h/\partial t) \to
b^{\alpha - z}\,(\partial h/\partial t)$, and the most relevant term of deterministic
part renormalizes as ${\cal G}(\nabla h) \to b^{\alpha \,m - (n + m)} {\cal G}(\nabla
h)$. In order to have scaling the three terms must be equally relevant and we find
\begin{equation}
\label{scal-rel-h-c}\left\{ \begin{array}{l}
      (m-1) \alpha + z -(m+n) = 0   \\
     2 \alpha - z + d = 0.
\end{array}
\right.
\end{equation}
These scaling relations are satisfied for any conserved model with
the appropriate (renormalized) values of $n$ and $m$. Regarding
the values of $m$ and $n$ the same comments as above apply here.

We can eliminate $z$ from Eq.(\ref{scal-rel-h-c}) to obtain $(m+1)
\alpha = m + n -d$. Equivalently, we have $(m+1)\hat{\alpha} =
n-1-d$ from Eq.(\ref{scal-rel-slope}). This allows us to eliminate
any dependence on the unknown exponents $n$ and $m$ and find the
relationships
\begin{equation}
\label{general}
     \alpha = \hat{\alpha} + 1 \:\:{\rm and}\:\:
     z = \hat{z},
\end{equation}
which are exact and valid for any growth model with conserved dynamics. The relations
(\ref{general}) immediately imply that for surface growth with conserved dynamics either
the scaling is Family-Vicsek ($\hat{\alpha} < 0$ and $\alpha < 1$), or it is super-rough
(if $\hat{\alpha} > 0$, then $\alpha > 1$). In the latter case, we can calculate the
local roughness exponent $\alpha_{loc}$ by making use of the scaling relation
$\alpha_{loc} = \alpha - z\kappa$ (valid whenever $\kappa,\hat{\alpha} > 0$) to obtain
that $\alpha_{loc} = \alpha - \hat{\alpha} = 1$, which shows that, if anomalous scaling
occurs in conserved growth, it can only be in the form of super-roughening.

\paragraph{Non-conserved dynamics.--} In this case
the lack of conservation leads to the renormalization of the noise
amplitude $D$ and the violation of the hyperscaling relation in
(\ref{scal-rel-h-c}). A one-loop dynamic renormalization group
calculation for the spectral function $\langle h(-{\bf k},t) h({\bf
k},t) \rangle$ yields the flow of the noise intensity $dD(l)/dl =
D\,[z-d-2\alpha+g(l)^2)]$, where $g(l)$ is the coupling constant,
which renormalization flow depends on the actual form of the most
relevant nonlinearity. This shows that in the scaling relations
(\ref{scal-rel-h-c}), only valid for conserved growth, we must
replace the hyperscaling relation by $z-d-2\alpha+g_*^2=0$, where
$g_*$ is the fixed point. This leads to the relation $\hat{\alpha} =
\alpha - 1 + g_*^2/(m+1)$ instead of (\ref{general}), which clearly
shows that the naive scaling $\nabla h \sim b^{\alpha-1}$ is
generally incorrect for non-conserved dynamics.

The scaling relations for the critical exponents of the surface slopes
(\ref{scal-rel-slope}), and in particular the condition for anomalous scaling
(\ref{AS-condit}), are also valid in the case of non-conserved growth. Therefore, we can
analyze the potential form of the nonlinear terms in Eq.(\ref{langevin}) that are
required for anomalous roughening in the case of non-conserved growth. In general, the
most relevant nonlinearity will be of the form ${\cal G}(\nabla h) \sim (\nabla^{n_1} h)
\cdots (\nabla^{n_N} h)$. However, not any term can actually appear in a leading order
expansion of ${\cal G}$. Under the assumption of {\em relaxational} dynamics
\cite{relax}, the shift invariance symmetry $h \to h + c$ greatly bounds the form of the
possible nonlinearities. This was studied in detail by Hentschel in Ref.\ \cite{hents},
where he showed that for growth models obeying non-conserved relaxational dynamics the
expansion of ${\cal G}$ to lowest order compatible with shift invariance is generically
given by
\begin{eqnarray}
\partial_t h=2\Gamma\,F^\prime\nabla^2h + 2\Gamma\,F^{\prime\prime}
\nabla[(\nabla h)^2]\cdot\nabla h -
\nonumber\\
- s\Gamma [F - 2F^\prime(\nabla h)^2] + \eta, \label{expan}
\end{eqnarray}
where $\Gamma$ and $F$ are functions of $(\nabla h)^2$, and {\em primes} denote
derivatives with respect to the argument \cite{hents}. Ulterior expansion of $\Gamma =
\Gamma_0 + \Gamma_1 (\nabla h)^2 + \cdots$ and $F = F_0 + F_1 (\nabla h)^2 + \cdots$ in
powers of $(\nabla h)^2$ provides all the terms compatible with shift invariance. In
general, all the possible nonlinear terms in the expansion (\ref{expan}) have the form
$\Upsilon^{2a}$, $(\nabla\cdot\Upsilon)\Upsilon^{2b}$ or $\Upsilon^{2c}\nabla
\Upsilon^2\cdot \Upsilon$. This means that, at any order in the expansion the only
allowed terms compatible with shift invariance lead to $n\leq 1$, and these values are
incompatible with the requirement Eq.\ (\ref{AS-condit}) in any dimension $d$. Since any
term compatible with shift invariance prevents condition (\ref{AS-condit}) from being
satisfied, we conclude that anomalous roughening (either intrinsic or super-rough) cannot
take place in non-conserved growth equations.

\paragraph{Conclusions.--} We have shown that
symmetries and conservation severely bound the form of the most
relevant terms that can exist in the long wavelength description of
growth models exhibiting anomalous scaling. This has been exploited
to show that intrinsic anomalous roughening cannot occur in local
growth models, but some conserved dynamics may display
super-roughening for some types of interaction terms. Our
conclusions are also valid in the case of growth driven by conserved
noise in Eq.(\ref{langevin}) after transforming $d \rightarrow d+2$
in the scaling relationships (7) and (10), which leads to the same
general conclusions. These results are useful in the search for
continuum models and the general problem of assigning universality
classes to experiments or discrete growth models. Our results imply
that disorder and/or non-local effects (like shadowing or bulk
diffusion) must be responsible for the occurrence of intrinsic
anomalous roughening in experimental systems
\cite{yang,jef,huo,jacobo,fracture,soriano}. Regarding simulations
of discrete models of local growth where anomalous scaling has been
reported  \cite{schro,kotrla}, like for instance in simulations of
the Wolf-Villain model \cite{wv} for ideal molecular-beam epitaxy,
our results imply that those anomalous scaling exponents have to be
taken with caution since they are effective exponents corresponding
to a transient, hence nonuniversal, regime. This conclusion is in
agreement with most recent simulation results of the Wolf-Villain
model in large system sizes and long times \cite{kotrla} and
theoretical arguments \cite{krug2}.

\begin{acknowledgements}
We wish to thank Miguel A. Rodr{\'\i}guez and Rodolfo Cuerno for
stimulating discussions. This work was supported by the CICyT
(Spain) through Grant No. BFM2003-07749-C05.

\end{acknowledgements}


\begin{thebibliography}{99}

\bibitem{krug-rev} J. Krug, Adv. Phys. {\bf 46}, 139 (1997).

\bibitem{fv} F.\ Family and T.\ Vicsek,
J.\ Phys.\ A {\bf 18}, L75 (1985).

\bibitem{krug} J. Krug,
Phys.\ Rev.\ Lett.\ {\bf 72}, 2907 (1994).

\bibitem{lopez97}
J.\ M.\ L\'opez, M.\ A.\ Rodr{\'\i}guez, and R.\ Cuerno, Phys.\
Rev.\ E {\bf 56}, 3993 (1997); Physica A {\bf 246}, 329 (1997).

\bibitem{groove} J.\ G.\ Amar, P.\ -M.\ Lam, and F.\ Family,
Phys.\ Rev.\ E {\bf 47}, 3242 (1993).

\bibitem{schro} M.\ Schroeder {\it et. al.},
Europhys.\ Lett.\ {\bf 24}, 563 (1993).

\bibitem{das} S.\ Das Sarma, S.\ V.\ Ghaisas, and
J.\ M.\ Kim, Phys.\ Rev.\ E {\bf 49}, 122 (1994); S.\ Das Sarma,
{\it et. al.}, Phys.\ Rev.\ E {\bf 53}, 359 (1996).

\bibitem{kotrla} P.\ Smilauer and M.\ Kotrla,
Phys.\ Rev.\ B {\bf 49}, R5769 (1994)

\bibitem{infrared} J.K. Bhattacharjee, S. Das Sarma and
R. Kotlyar, Phys.\ Rev.\ E {\bf 53}, R1313 (1996).

\bibitem{ryu} C.\ S.\ Ryu, K.\ P.\ Heo, and I.\ Kim, Phys.\ Rev.\
E {\bf 54}, 284 (1996).

\bibitem{dasgupta} C. Dasgupta, S. Das Sarma and J.M. Kim,
Phys.\ Rev.\ E {\bf 54}, R4552 (1996).

\bibitem{lopez96} J.M. L\'opez and  M.A. Rodr{\'\i}guez,
Phys. Rev. E {\bf 54}, R2189 (1996).

\bibitem{ala} J.\ Asikainen {\it et. al.}, Phys.\ Rev.\ E {\bf
65}, 052104 (2002).

\bibitem{mario} M.\ Castro, {\it et. al.}
Phys.\ Rev.\ E {\bf 57}, R2491 (1998).

\bibitem{cvd} R.\ Cuerno and M.\ Castro,
Phys.\ Rev.\ Lett.\ {\bf 87}, 236103 (2001)


\bibitem{yang} H.\ -N.\ Yang, G.\ -C.\ Wang and T.\ -M.\ Lu,
Phys.\ Rev.\ Lett.\ {\bf 73}, 2348 (1994).

\bibitem{jef} J.\ H.\ Jeffries, J.\ -K.\ Zuo, and M.\ M.\ Craig,
Phys.\ Rev.\ Lett.\ {\bf 76}, 4931 (1996).

\bibitem{huo} S.\ Huo and W.\ Schwarzacher,
Phys.\ Rev.\ Lett.\ {\bf 86}, 256 (2001).

\bibitem{jacobo} J.\ Santamar{\'\i}a {\it et. al.},
Phys.\ Rev.\ Lett.\ {\bf 89}, 190601 (2002).

\bibitem{fracture} J.\ M.\ L\'opez and J.\ Schmittbuhl,
Phys.\ Rev.\ E {\bf 57}, 6405 (1998); S.\ Morel {\it et. al.},
Phys.\ Rev.\ E {\bf 58}, 6999 (1998).

\bibitem{soriano} J.\ Soriano {\it et. al.},
Phys.\ Rev.\ Lett.\ {\bf 89}, 026102 (2002).

\bibitem{bru} A.\ Bru {\it et. al.}
Phys.\ Rev.\ Lett.\ {\bf 81}, 4008 (1998).

\bibitem{lopez99} J.\ M.\ L\'opez,
Phys.\ Rev.\ Lett.\ {\bf 83}, 4594 (1999).

\bibitem{ramasco} J.\ J.\ Ramasco, J.\ M.\ L\'opez,
M.\ A.\ Rodr{\'\i}guez, Phys. Rev. Lett. {\bf 84}, 2199 (2000).

\bibitem{barabasi} A.\ -L.\ Barab\'asi
and H.\  E.\ Stanley, {\em Fractal Concepts in Surface Growth}
(Cambridge University Press, Cambridge, 1995).

\bibitem{4th_order} S.\ Das Sarma and R.\ Kotlyar, Phys.\ Rev.\ E
{\bf 50}, R4275 (1994).

\bibitem{sgg} T.\ Sun, H.\ Guo, and M.\ Grant,
Phys.\ Rev.\ A {\bf 40}, R6763 (1989).

\bibitem{flory} H.\ G.\ E.\ Hentschel and F.\ Family,
Phys.\ Rev.\ Lett.\ {\bf 66}, 1982 (1991).



\bibitem{wv} D.\ E.\ Wolf and J.\ Villain,
Europhys.\ Lett.\ {\bf 13}, 389 (1990).


\bibitem{hents} H.\ G.\ E.\ Hentschel, J.\ Phys.\ A {\bf 27}, 2269
(1994).

\bibitem{relax} Relaxational dynamics is a reasonable assumption for surface growth models
and corresponds to model A dynamics of P.\ C.\ Hohenberg and B.\ Halperin Rev.\ Mod.\
Phys.\ {\bf 49}, 435 (1977)

\bibitem{krug2} J.\ Krug, M.\ Plischke, and M.\ Siegert, Phys.\
Rev.\ Lett.\ {\bf 70}, 3271 (1993).

\end{thebibliography}
\end{document}